\documentclass{jaa}
\usepackage{natbib}
\usepackage{graphicx}
\usepackage{xcolor}
\usepackage{soul}
\graphicspath{{./}{figures/}}

\begin{document}\sloppy

\title{Structural Analysis of Open Cluster Bochum 2\\ }


\author{Harmeen Kaur\textsuperscript{1,*}, Saurabh Sharma\textsuperscript{2}, Alok Durgapal\textsuperscript{1}, Lokesh K Dewangan\textsuperscript{3}, Aayushi Verma\textsuperscript{2}, Neelam Panwar\textsuperscript{2}, Rakesh Pandey\textsuperscript{3} and Arpan Ghosh\textsuperscript{2}}
\affilOne{\textsuperscript{1} Center of Advanced Study, Department of Physics, DSB Campus, Kumaun University Nainital, 263002, India.\\}
\affilTwo{\textsuperscript{2} Aryabhatta Research Institute of Observational Sciences (ARIES), Manora Peak, Nainital 263002, India.\\}
\affilThree{\textsuperscript{3} Physical Research Laboratory, Navrangpura, Ahmedabad, 380009, India\\}

\twocolumn[{

\maketitle

\corres{harmeenkaur.kaur229@gmail.com}

\msinfo{7 February 2023}{28 February 2023}

\begin{abstract} 

    We present the results from our deep optical photometric observations of  Bochum 2 (Boc2) star cluster obtained using the $1.3$m Devasthal Fast Optical Telescope along with archival photometric data from Pan-STARRS2/2MASS/UKIDSS surveys. We also used high-quality parallax and proper motion data from the $Gaia$ Data Release 3. We found that the Boc2 cluster has a small size ($\sim$1.1 pc) and circular morphology. 
    Using $Gaia$ parallax of member stars and isochrone fitting method, the distance of this cluster is estimated as  $3.8\pm0.4$ kpc.
     We have found that this cluster holds young ($\sim5$ Myr) and massive (O$7-$O$9$) stars as well as an older population of low mass stars.
     We found that the massive stars have formed in the inner region of the Boc2 cluster in a recent epoch of star formation.
    We have derived mass function slope ($\Gamma$) in the cluster region as 
     $-2.42\pm0.13$ in the mass range $\sim0.72<$M/M$_{\odot}<2.8$. 
    The tidal radius of the Boc2  cluster  ($\sim7-9$) is much more than its observed radius ($\sim1.1$ pc). This suggests that most of the low-mass stars in this cluster are the remains of an older population of stars formed via an earlier epoch of star formation.

\end{abstract}


}]


\doinum{12.3456/s78910-011-012-3}
\artcitid{\#\#\#\#}
\volnum{000}
\year{0000}
\pgrange{1--}
\setcounter{page}{1}
\lp{1}

%
%
%
\section{Introduction}

    Young open clusters are distinctive to understand the 
    process of star formation and stellar evolution. The young star clusters contain both high-mass and low-mass stars.
     The physical properties of stars along with the distribution of ionized gas, dust (cold as well as warm), and  molecular gas can give us observational hints about the physical processes that conduct their formation and evolution \citep{2008MNRAS.384.1675J}. 
     Many star cluster also show the distribution of massive stars towards their central region. Whether this segregation of  massive stars occurs due to an evolutionary effect or is of primordial origin is vague \citep{2006AJ....131.1574L}.
   
    For the present study, we have selected Bochum 2 young open cluster, hereafter Boc2, centered at $\alpha_{J2000}$: 06$^h$48$^m$51.6$^s$, $\delta_{J2000}$: +00$^\circ$23$^\prime$32.719$^\prime$$^\prime$.
    This cluster is located in Galactic plane towards the $3^{rd}$ quadrant ($l=212^\circ.28878, b=-0^\circ.39462$) in the far northern outskirts of an H\,{\sc ii} region Sh $2-284$.
    \citet{2007A&A...470..161R} have proposed that
    these  regions are part of a star forming region located in the Milky Way's Norma (Outer) arm at a distance of $7.9\pm0.3$
    kpc. 
    During a search through the catalogue of luminous stars \citet{1975A&AS...20...85M} identified Boc2 cluster, that accommodate O-type stars. \cite{1979A&AS...38..197M} provided information about the MK spectral types of the three brightest member stars of the Boc2 cluster as O9V, O7V and O9V.  They also spectroscopically derived mean reddening of 
    $E(B-V)=0.84$  mag and a distance of $4.8$ kpc for this cluster. 
    Later, \cite{1993AJ....105.1831T} found differential reddening 
    within the cluster and reported the distance, reddening and age of 
    this cluster as $\sim$5.5 kpc, $E(B-V)=0.89$ mag and $5$ Myr, respectively.
    \cite{1995MNRAS.277.1269M} undertook spectro-photomeric studies of this 
    cluster and estimated the distance, mean reddening and age of the
    cluster as $\sim$6 kpc, $E(B-V)= 0.80$ and $7$ Myr, respectively. They 
    also confirmed the spectral type of brightest stars, earlier derived 
    by \citet{1979A&AS...38..197M}. 
    Recently, \cite{2016A&A...585A.101K} studied the Galactic star cluster 
    population based on the Milky Way Star Cluster (MWSC) survey. To
    determine cluster parameters and membership they used a 
    combination of uniform kinematic and NIR 
    photometric data gathered from the all-sky catalogue PPMXL 
    \citep{2010AJ....139.2440R}. The study estimated distance, extinction and
    age of this cluster as 2.8 kpc, $E(B-V)=0.87$ mag and $4.6$ Myr. 

\begin{figure}
\centering
\includegraphics[width=0.48\textwidth]{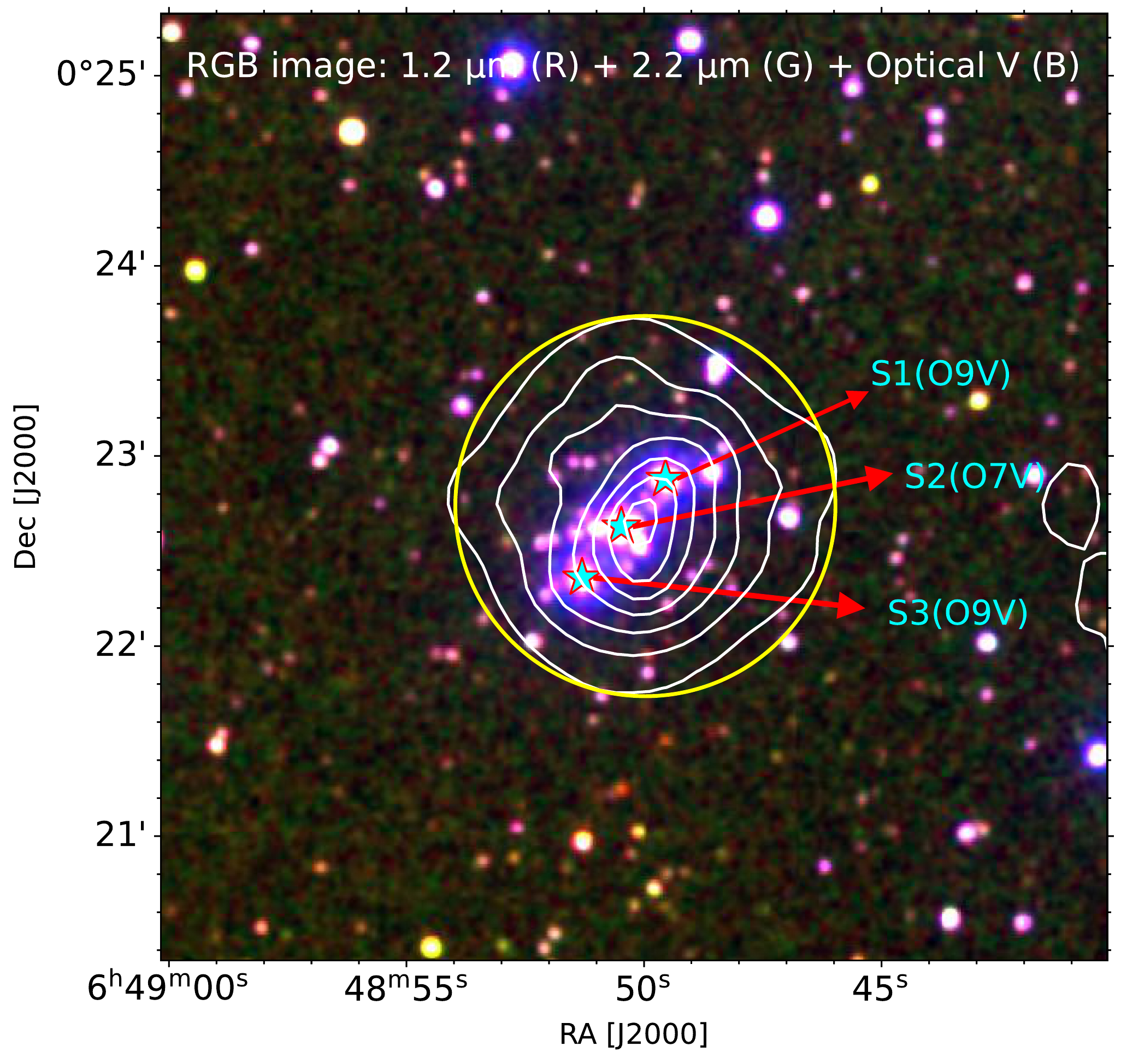}
\caption{Color-composite image of Boc2 cluster region (Red: 2MASS J band (1.2 $\mu$m); Green: 2MASS K band (2.2 $\mu$m) and  Blue: 
Optical V band, taken from 1.3m DFOT) overlaid with the white isodensity contours generated from the NIR catalog (2MASS+UKIDSS; cf. Section 3.1). The lowest level for the isodensity contours is 3$\sigma$ above the mean stellar density (i.e., 0.29 stars arcmin$^{-2}$) with a step size of 1$\sigma$ (0.1 stars arcmin$^{-2}$). Yellow circle  
encloses the lowest density contour which delineates cluster's region having a radius 1$^\prime$. Three massive stars located inside the cluster region 
are also marked with star symbols along 
with their spectral type (cf. Section 3.3)}.\label{zrgb}

\end{figure}

   All the previous studies on Boc2 were based on the shallow optical 
    data and not so precise membership estimation of the cluster members were done. Therefore, with the availability of high quality proper motion data from Gaia DR3 \citep{2016A&A...595A...1G, 2018A&A...616A...1G} along with new deep and wide-ﬁeld multi band (optical-to-near-infrared) data sets (cf. Section 2), 
    we have revisited the Boc2 cluster to study in detail about the formation and evolution processes of stars. Also, the extracted parameters of this cluster can be utilized to enrich the sample
    of clusters required to study Galactic structure and dynamics.

    The structure of this paper is as follows. In Section 2, a brief
    description of observation and data reduction, and the details
    of available data sets from various archives is presented. Section 3, describes result and analysis of this study which includes the study of the structure of this cluster, membership probability, and the estimation of basic parameters of the cluster (i.e., extinction, distance, age, mass function etc.). In Section 4, we discussed about the results and concluded them in Section 5.



\section{Observation and Data Reduction}

\subsection{Optical data set}
 The optical CCD $UBV{I}_c$  photometric data of the Boc2 cluster,
        centered at $\alpha_{J2000}$: 06$^h$48$^m$51.6$^s$, $\delta_{J2000}$: 
    	+00$^\circ$23$^\prime$32.719$^\prime$$^\prime$,
        were acquired by using the $2048\times 2048$ pixel$^2$  CCD mounted on F/4 cassegrain focus of Devasthal Fast Optical Telescope
        (DFOT) ARIES, Nainital, India.
        The entire chip covers a field-of-view (FOV) of 
        $\sim18.5\times18.5$ arcmin$^2$ (Plate scale: 0.54
        arcsec/pixel).
        To improve the signal to noise ratio (SNR), the observations were carried out in the binning mode of $2\times2$ pixels.
        The read-out noise and gain of the CCD are 8.29 $e^-$ and 2.2 $e^-$/ADU
        respectively. The average FWHM (stellar profile) of the stars 
        were $\sim2.2$ arcsec.
        The broad-band $UBV{I}_c$ observations of the Boc2 cluster were
        standardized by observing stars in the
        SA98 field \citep{1992AJ....104..340L}.
        A number of bias frames and twilight-flat frames were also taken during
        observations. Short and deep (long) exposure frames
        were taken to observe both bright and faint stars in the field.
        The complete log of the observations is given in Table ~\ref{log}.

The CCD data frames were reduced by using the computing facilities available at the  Center of Advanced Study, Department of Physics, Kumaun University, 
	Nainital and ARIES, Nainital. Initial processing of the data frames were done 
    by using the IRAF\footnote{IRAF is distributed by National Optical Astronomy
    Observatories, USA.} and ESO-MIDAS\footnote{ ESO-MIDAS is developed  
    and maintained by the European Southern Observatory.} data reduction packages. 
    Photometry of the cleaned frames were carried out by using DAOPHOT-II
    software \citep{1987PASP...99..191S}. The point spread function (PSF) was
    obtained for each frame by using several uncontaminated stars. 
    We used the DAOGROW program for construction of an aperture
    growth curve required for determining the difference between the aperture and
    profile-fitting magnitudes. Calibration of the instrumental magnitudes to 
    the standard system was done by using the procedures outlined by \citet{1992ASPC...25..297S}. The total 40 standard stars were used in the photometric calibration of this cluster.
    The calibration equations derived by the least-squares linear regression are 
    as follows:

        \begin{equation}
        \begin{split}
        u&= U + (5.380\pm0.007)\\ 
        &-(0.054\pm0.008)(U-B) + (0.433\pm0.009)X_U,
        \end{split}
        \end{equation}

        \begin{equation}
        \begin{split}
        b&= B + (4.537\pm0.015)\\
        &-(0.192\pm0.011)(B-V) + (0.215\pm0.015)X_B,
        \end{split}
        \end{equation}
        
        \begin{equation}
        \begin{split}
        v&= V + (3.296\pm0.009)\\
        &+(0.009\pm0.006)(V-I_c) + (0.101\pm0.008)X_V,
        \end{split}
        \end{equation}

        and 
        
        \begin{equation}
        \begin{split}
        i_c&= I_c + (3.284\pm0.005)\\
        &+(0.000\pm0.002)(V-I_c) + (0.039\pm0.006)X_I. 
        \end{split}
        \end {equation}

	 \noindent
        where $U,B,V $ and $I_c$ are the standard magnitudes
        and $u,b,v$ and $i_c$ are the instrumental aperture magnitudes,
        which are normalized per second of exposure time and $X_{U}$, $X_{B}$, $X_{V}$ and $X_{I}$ are the airmass in respective filters. 
                
\begin{figure}
    \centering
    \includegraphics[width=.45\textwidth]{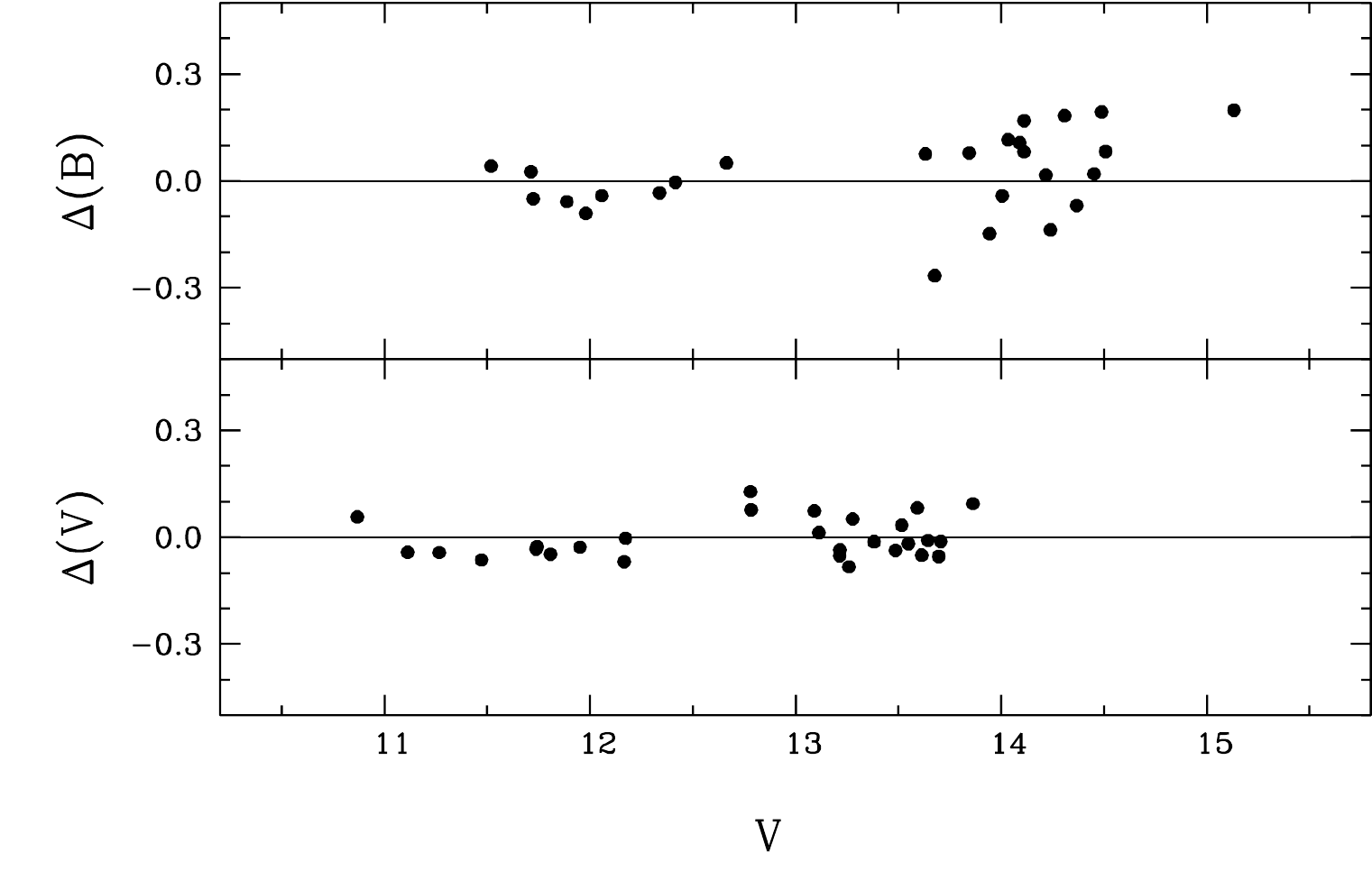}
    \caption{Comparison of present photometric data with available apass data as a function of present V mag.}\label{diff}
    \end{figure}

\begin{figure}
    \centering
    \includegraphics[width=.48\textwidth]{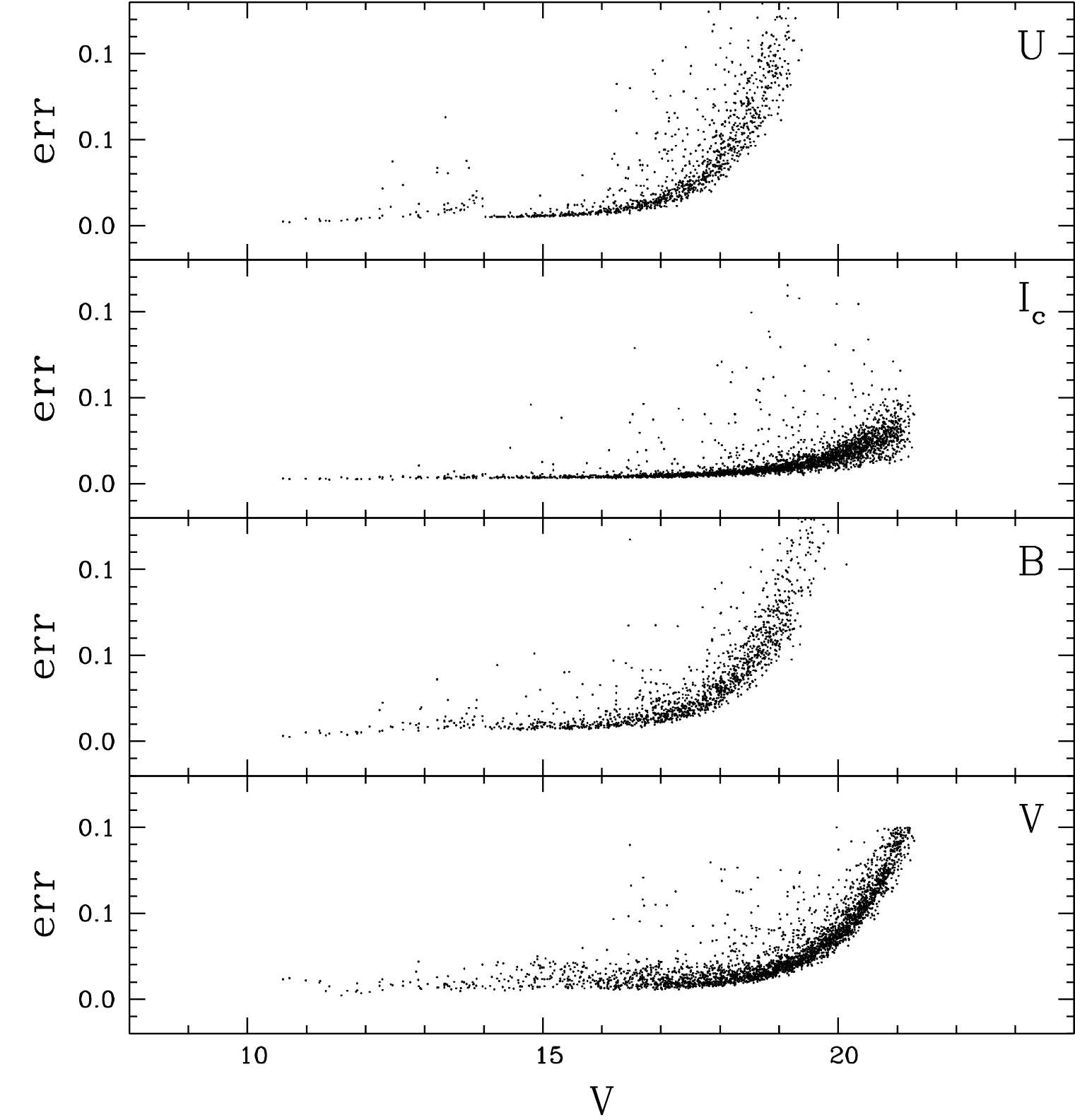}
    \caption{ $V,B,I_{c},U$ photometic errors as a function of V magnitude.}\label{err}
    \end{figure}

        We have also carried out a comparison of the present photometric 
        data with available APASS data. The difference $\Delta$(present$-$apass) as a function of $V$ magnitude is shown in Figure~\ref{diff}. The comparison indicates that the magnitudes obtained in the present work are in fair agreement with those available in the literature.
        The distinctive  DAOPHOT errors in different bands as a function of V magnitude shown in Figure~\ref{err}. In this bright and faint stars are taken from  short and long  frames, respectively. 
        It can be seen that at fainter magnitude limit errors become large ($>$0.1) and were not used in present work. In this study,
        a total of 2824 sources have been identiﬁed with detection at least in the $V$ and $I_{c}$ bands and having photometric errors less than 0.1 mag up to $V\sim21.3$ mag.

\subsection{Archival data sets}

   We used the near-infrared (NIR) point source catalog\footnote{https://irsa.ipac.caltech.edu/Missions} and images\footnote{https://skyview.gsfc.nasa.gov/current/cgi/query.pl} from 2MASS \citep{2003yCat.2246....0C, 2006AJ....131.1163S} and UKIDSS\footnote{http://wsa.roe.ac.uk}
    \citep{2007MNRAS.379.1599L}.  We have also used the Panoramic Survey Telescope and Rapid Response System\footnote{https://catalogs.mast.stsci.edu/} (PanSTARRS1 or PS1) data release 2 \citep{2016arXiv161205560C} 
    and recently available high quality proper motion data from 
    Gaia\footnote{https://gea.esac.esa.int/archive/} Data Release 3 \citep{gaia1234,gaia5678}.
    For our photometric analysis, we used only those sources which have uncertainties less than 0.1 mag.

\begin{table}
\tabularfont
\caption{Log of observations}\label{log}
\begin{tabular}{lccc}
\topline
Date of observations/Filter& Exp.(sec)$\times$ No. of frames\\\midline
	&SA98\\
	10 January 2013\\
	$U$   &  $60\times1,90\times3, 120\times2$\\
	$B$   &  $30\times1,40\times3,60\times2$\\
	$V$   &  $30\times1,40\times5$\\
	$I_c$ &  $10\times1,20\times5$\\
	\\
    & Bochum 2\\
    10 January 2013\\
    $U$   &  $20\times3,600\times3$\\
	$B$   &  $10\times3,600\times3$\\
	$V$   &  $10\times3,600\times3$\\
	$I_c$ &  $10\times3,600\times4$\\
 
\hline
\end{tabular}
\end{table}

\section{Results and Analysis}

\subsection{Structure of the Bochum 2 cluster}

    As literature put forward, this cluster is young and have distribution of gas and dust around it. We have used the stellar number density
    profile, obtained from the NIR catalog of stars, to study the structure of this cluster (cf. Figure~\ref{zrgb}). 
    The NIR catalog is compiled from the 2MASS (for stars with $J_{mag}<13$; bright stars) and UKIDSS (for stars  with $J_{mag}\geq13$; faint stars) surveys covering $18^\prime.5 \times18^\prime.5$ 
    FOV (similar to our optical observations) around the cluster region.
    The stellar number density maps were generated using the nearest neighbour (NN) method as described by \citet{2005ApJ...632..397G} and
    \citet{2020MNRAS.498.2309S}. We took the radial distance necessary to encompass the $20^{th}$ nearest stars and computed the local surface density in a grid size of $6$ arcsec \citep{2009ApJS..184...18G}.
    Figure~\ref{zrgb}, constitutes the color image of 
    $5^\prime \times 5^\prime$ area around the Boc2 cluster region, composed with
     Red: 2MASS J band ($1.2\mu$m); Green: 2MASS K band ($2.2 \mu$m)
     and Blue: Optical V band, taken from 1.3m DFOT.
    The stellar number density contours derived by above method are 
    superimposed in Figure~\ref{zrgb} as white contours smoothened to a grid of size $3\times3$ pixels. The lowest contour is $3\sigma$ above the mean of stellar density ($0.29$ star arcmin$^{-2}$) with a step size of $1\sigma$
     ($0.1$ stars arcmin$^{-2}$). Although the bright stars in the cluster region suggests an elongated morphology, whereas the cluster obtained from deep NIR isodensity contours show almost circular morphology. The yellow circle in Figure~\ref{zrgb} encloses the lowest density contour and delineates cluster
     region with a radius of 1 arcmin centered at $\alpha_{J2000}$: 06$^h$48$^m$50.17$^s$, $\delta_{J2000}$: +00$^\circ$22$^\prime$42.17$^\prime$$^\prime$. Inside the cluster region (lowest density contour or yellow circle of
     radius 1$^\prime$) three massive stars are also marked as star symbols in Figure~\ref{zrgb}. These stars are named as S1, S2 and S3 with spectral type O9V, O7V and O9V, respectively \citep[for details refer,][]{1995MNRAS.277.1269M}.

    \begin{figure*}
    \centering\includegraphics[height=0.35\textheight]{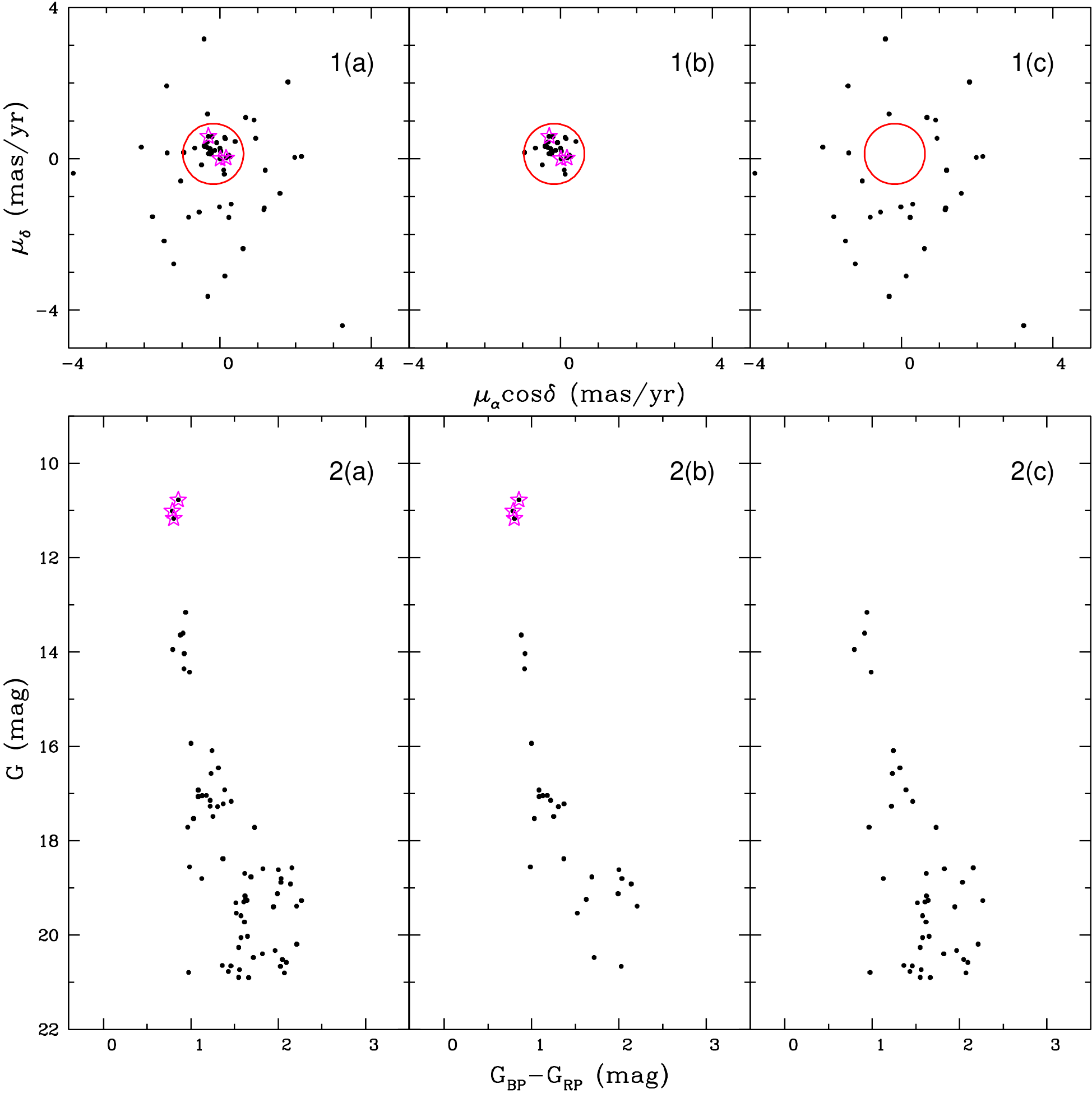}
    \centering\includegraphics[height=0.35\textheight]{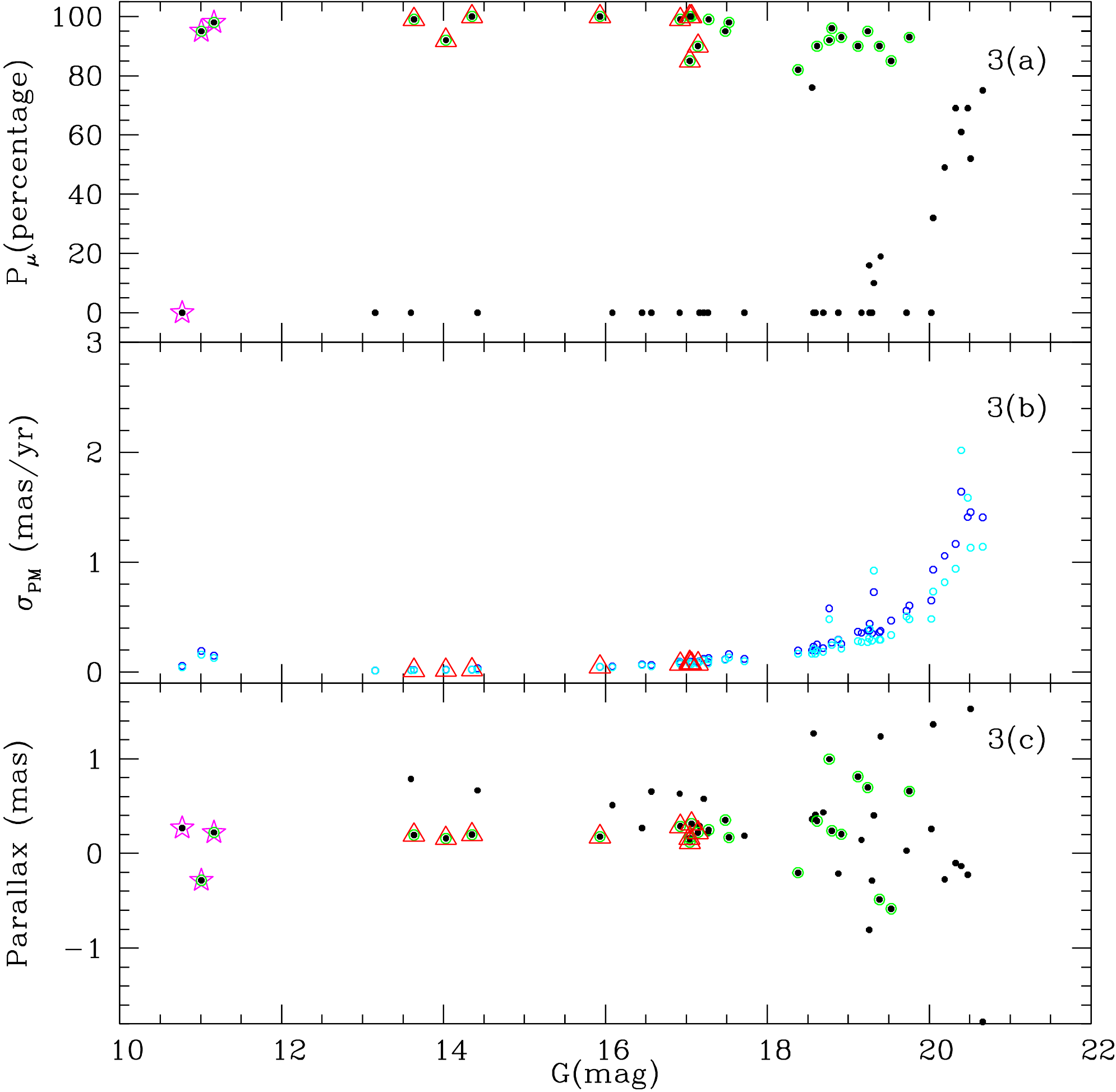}
    \caption{{\it Panel-1} shows the PM vector-point diagrams (VPDs) and {\it panel-2} shows $G_{mag}$
    vs. $G_{BP}-G_{RP}$ CMDs. 
    The left sub-panels [1(a) and 2(a)] show all stars located within 
    Boc2 cluster region (cf. Section 3.1), while the middle [1(b) and 2(b)] and right sub-panels [1(c) and 2(c)] show the probable cluster members and field stars. {\it Panel-3} depicts the  membership probability ($P_{\mu}$), PM 
    errors ($\sigma_{PM}$) and parallax of stars in the cluster region as a function of G magnitude. The probable member stars ($P_{\mu}> 80$) are shown by green
    circles while the 9 members used for distance estimation of the Boc2 cluster are
    depicted by red triangles (see text for details). Location of massive stars (O-type) can also be seen in all panels by star symbols.}\label{plx}
    \end{figure*}

    \begin{figure*}
    \centering\includegraphics[height=0.35\textheight]{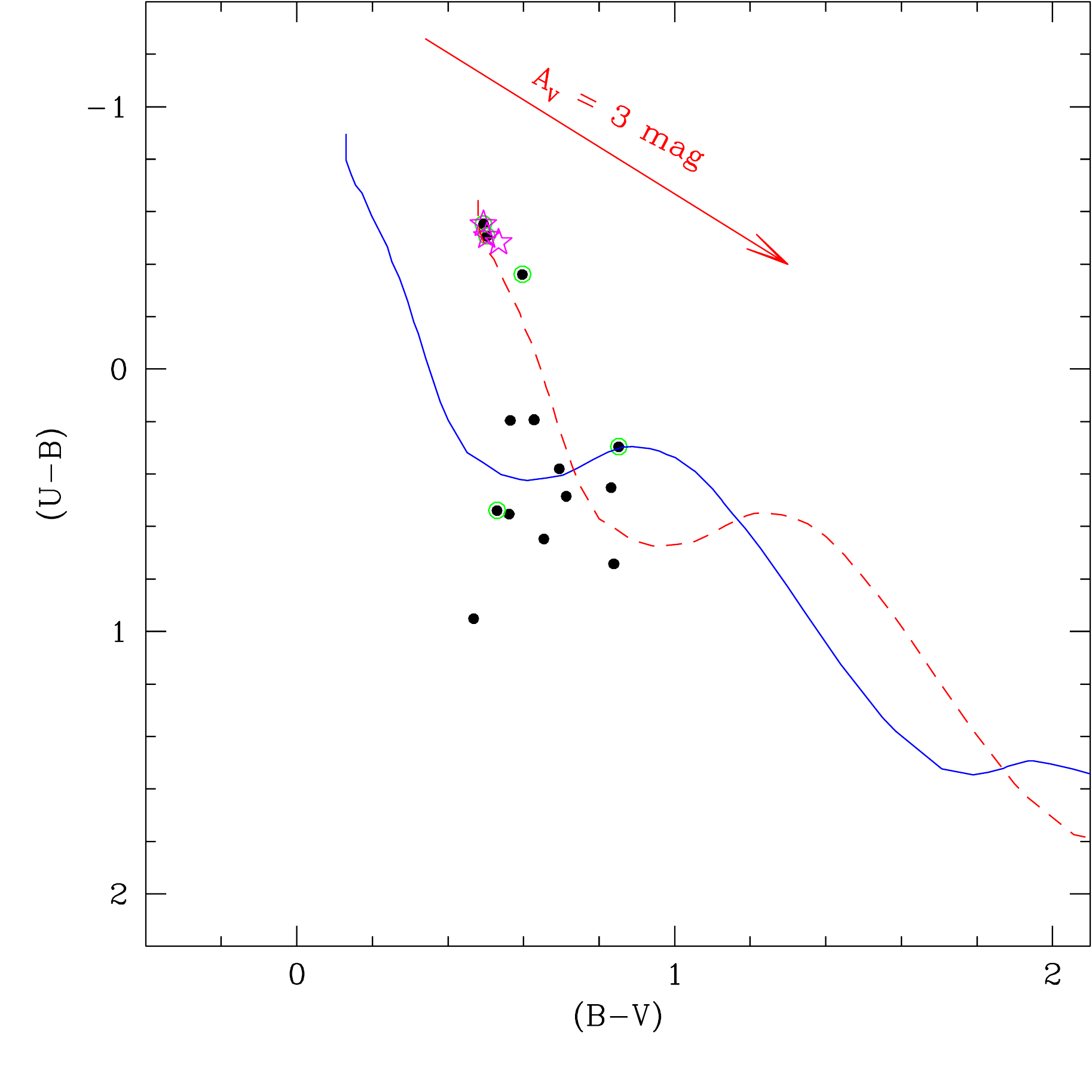}
    \centering\includegraphics[height=0.35\textheight]{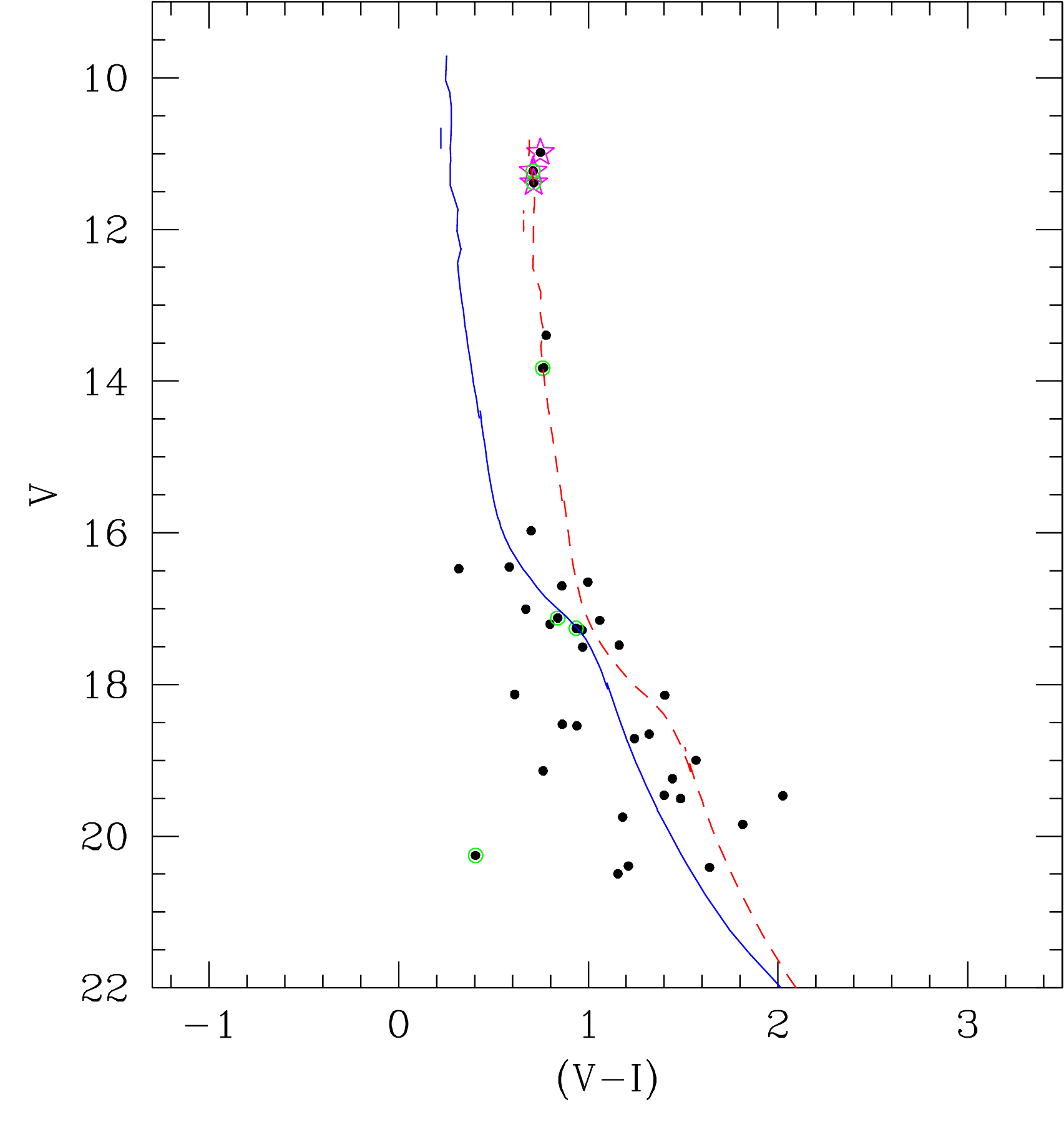}
    \caption{{\it Left panel}: $(U-B)$ vs.$(B-V)$ TCD for optically detected sources in the Boc2 cluster region. The  blue curve describes the
    intrinsic Zero Age Main Sequence (ZAMS) for $Z = 0.02$ by \cite{2013ApJS..208....9P} shifted along the reddening vector for $E(B-V)=0.45$ mag. The dotted red curve indicates same ZAMS but for $E(B-V)=0.80$ mag.
    Green open circles and magenta star symbols are the location of cluster members and massive O-type stars, respectively.
    {\it Right panel}: $V$ vs. $(V-I_{c})$ CMD for the same sources as described in the left panel. 
    The curves shows the ZAMS from \citet{2019MNRAS.485.5666P} corrected for a distance of $3.8$ kpc, and reddening $E(B-V) = 0.45$ mag (blue curve) and $E(B-V) = 0.80$ mag (red dotted curve), respectively, for the foreground and cluster reddening values, respectively. 
    }\label{cc}
    \end{figure*}

    \begin{figure*}
    \centering\includegraphics[height=0.4\textheight]{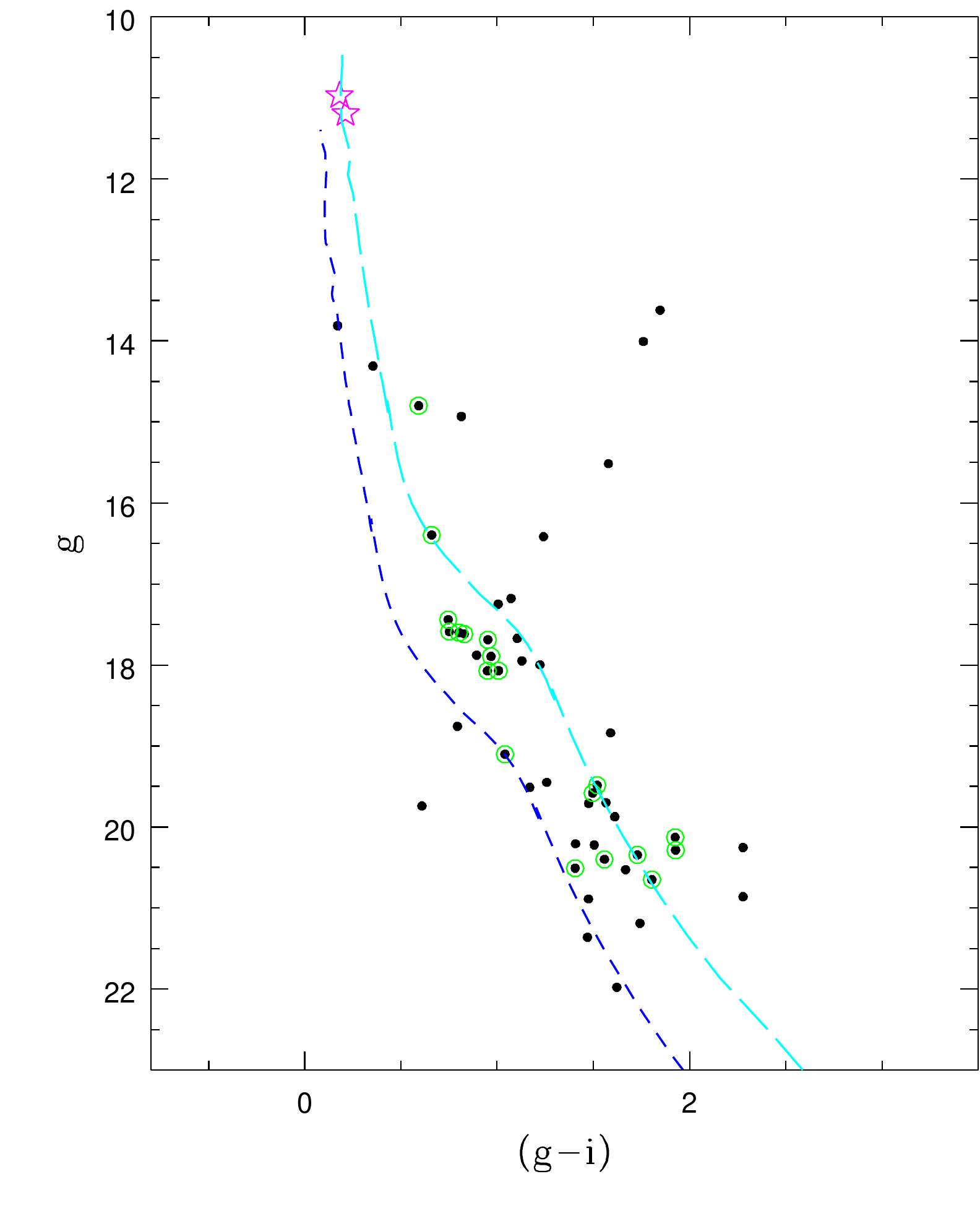}
    \centering\includegraphics[height=0.4\textheight]{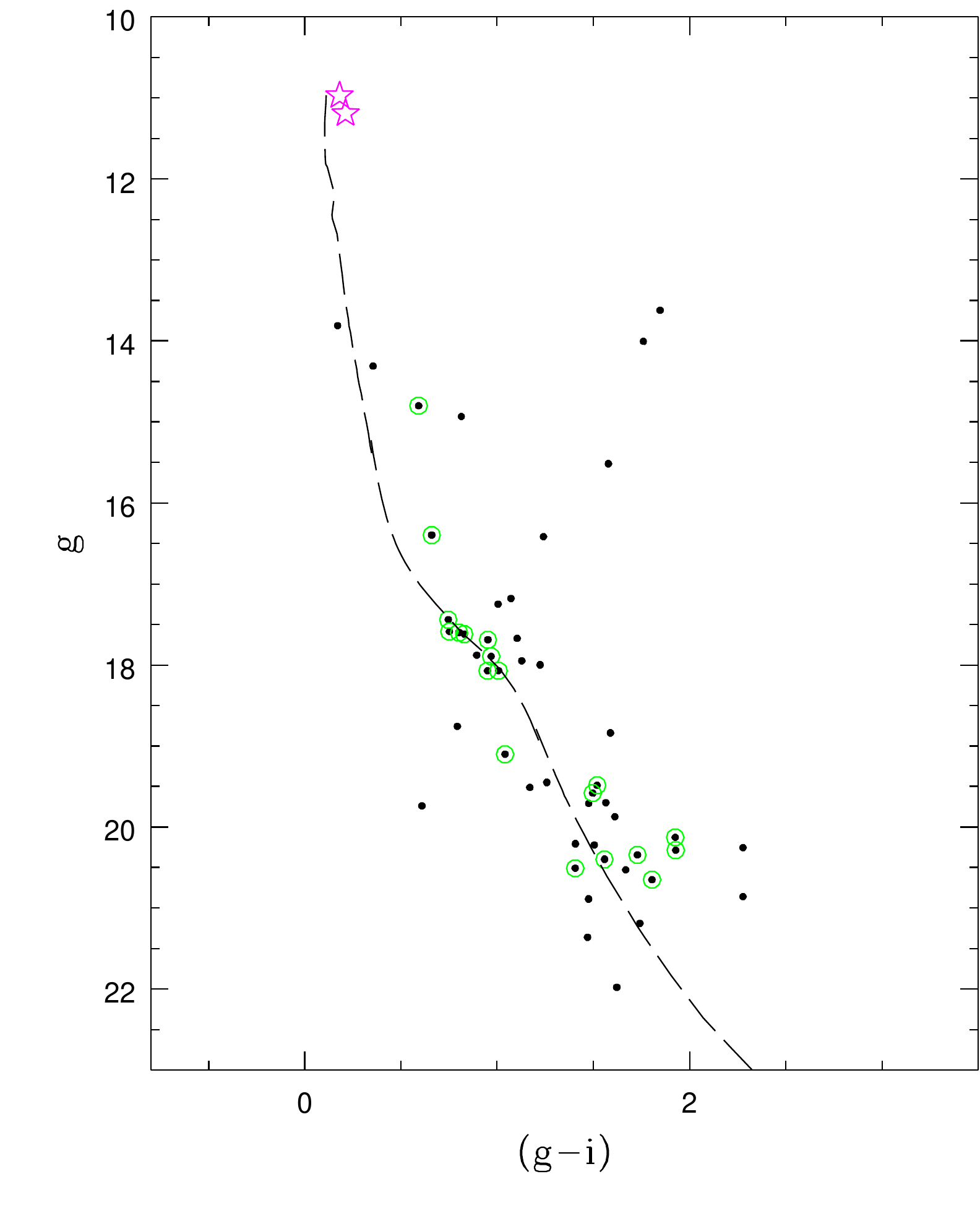}
    \caption{{\it Left panel:} PS1 $g$ vs. $(g-i)$ CMD for the stars in the cluster 
    region (black dots). Green open circles and magenta star symbols are the location of cluster members and massive O-type stars, respectively.
    The curves denote the ZAMS derived from \cite{2019MNRAS.485.5666P} corrected for 
    extinction and distance values reported in literature, i.e., 
    $A_{V}=2.76$ mag, distance $= 5.5$ kpc \cite{1975A&AS...20...85M} (blue dashed curve) and $A_{V}=2.70$ mag, distance $= 2.8$ kpc \cite{2016A&A...585A.101K} (cyan dashed curve). {\it Right panel}: Same as left panel but with ZAMS corrected for a distance of $3.8$ kpc and reddening of $E(B-V)=0.80$ mag (present work, black dashed curve).}\label{cmd}
    \end{figure*}

\subsection{Membership Probability}
    The astrometric selection of the cluster's members in large scale
     was always limited by the errors of the proper motions given by
     the catalogues before the $Gaia$ mission. Better membership
        determination made possible by the new generation of high precision
        astrometric $Gaia$ DR3 catalogue upto very faint limits \footnote{https://gea.esac.esa.int/archive/} (\citep{2016A&A...595A...1G, 2018A&A...616A...1G}. To determine the membership probability using $Gaia$ DR3, we adopted the 
    method described in \citet{1998A&AS..133..387B}.  This method is
    efficiently used  before in many studies \citep{2020MNRAS.498.2309S, 2020ApJ...891...81P, 2020ApJ...896...29K}. In this method, we first construct the frequency distributions of cluster 
    stars $(\phi^\nu_c)$  and field stars $(\phi^\nu_f)$ using the following equations given in \citet{2013MNRAS.430.3350Y}. 
    The $Gaia$ proper motion (PM) data of the stars located within the Boc2 cluster region is used to determine their membership 
    probability. 
    The PMs, $\mu_\alpha$cos($\delta$) and $\mu_\delta$, are plotted as vector-point diagrams (VPDs) in the panel 1 of Figure \ref{plx}. The panel 2 shows the corresponding $G$ versus $G_{BP} - G_{RP}$ Gaia color-magnitude diagrams (CMDs).
    The tight clump centering at 
     $\mu_{xc} = -0.175$ mas yr$^{-1}$, $\mu_{yc} = 0.133$ mas yr$^{-1}$,  having radius of $0.8$ mas yr$^{-1}$ in the VPD represents the probable cluster stars, and the remaining represent the distribution of the probable field stars.
    The left sub-panels show all stars, while the middle and right 
    sub-panels show the probable cluster members and field stars. 
    Assuming a distance of $3.8$ kpc (estimated in present study cf. Section 3.3) and a
    radial velocity dispersion of 1 kms$^{-1}$ for open clusters \citep{1989AJ.....98..227G}, the expected dispersion ($\sigma_c$) in PMs of the cluster would be $\sim0.05$ mas yr$^{-1}$.
    From the distribution of probable field stars,
    we have calculated, $\mu_{xf}=-0.26$ mas yr$^{-1}$, $\mu_{yf}=-0.55$ mas yr$^{-1}$,
    $\sigma_{xf}=1.90$ mas yr$^{-1}$ and $\sigma_{yf}=1.90$ mas yr$^{-1}$.
    These values are further used to construct the frequency distribution of cluster stars ($\phi_c^{\nu}$) and field stars ($\phi_f^{\nu}$) by using the equations given in \citet{2013MNRAS.430.3350Y} and then the value of membership probability of the cluster is estimated by using the following equation:
        
         \begin{equation}
        P_\mu(i) = {{n_c\times\phi^\nu_c(i)}\over{n_c\times\phi^\nu_c(i)+n_f\times\phi^\nu_f(i)}}
        \end{equation}
        
	   where $n_c$ ($=0.48$) and $n_f$($=0.52$) are the normalized numbers of stars for the cluster and field region ($n_c$+$n_f = 1$). The 
        membership probability (estimated above), errors in the PM, and parallax values are plotted as a function of G  magnitude in panel 3 of Figure~\ref{plx}.
     As can be seen in this plots, a high
     membership probability (P$\mu > 80 \%$) extends down to G $\sim20$ mag. At brighter magnitudes, there is a clear separation between cluster members and  field stars that reinforce this technique. Errors in PMs become very high at faint limits, and the maximum probability gradually decreases at those levels. 
     Therefore, on the basis of above analysis, 24 stars were assigned as cluster members of Boc2 cluster based on their high membership probability P$\mu>80\%$ (green circles with black rings in Figure~\ref{plx}).
     Location of massive stars (O-type; cf. Section 3.1)
     are also shown in all panels by the magenta star symbol.
     
\subsection{Reddening, Distance and Age of the cluster}

    The reddening in the direction of a cluster can be derived quite accurately by using
    the two-color diagrams (TCDs) \citep[cf.,][]{1994ApJS...90...31P,2006AJ....132.1669S}.
    In the left panel of Figure~\ref{cc},
    we show  the ($U-B$) versus ($B-V$) TCD of the stars located inside the Boc2 cluster. 
    The identified member stars  (green circles) from the Gaia data and most massive stars  (magenta star symbols) located inside the cluster region are also marked in the ﬁgure.
    To estimate the reddening in the cluster direction, the intrinsic zero-age main sequence \citep[ZAMS,][for $Z=0.02$]{2013ApJS..208....9P} is shifted along the reddening vector ($E(B-V)/E(U-B) = 0.72$), so that it matches with the distribution of member stars. 
    The amount of shift in the X-axis give a reddening ,$E(B-V)$, value of the cluster.
    The blue continuous curve and the red dotted curve are the ZAMS which are visually fitted to estimate the reddening value for the foreground stars ($E(B-V)_{f} = 0.45$ mag) and the massive stars ($E(B-V)_{cl} = 0.80$ mag), respectively. As the massive stars are members of the Boc2 cluster, we designated their reddening as cluster reddening value.
      
    As already discussed in Section 1, there is a large discrepancy in the earlier reported values of the distance of Boc2 cluster, i.e., 2.8 to 6 kpc. We have used the $Gaia$ data to revise the distance estimate of this cluster. For this purpose, we have selected $9$ 
    cluster members having  parallax error $<0.1$ (red triangles in Figure~\ref{plx}) and calculated the mean of the photo-geometric distances as reported by \citet{2021AJ....161..147B}. Thus, the distance of Boc2 cluster comes out to be $\sim 3.8\pm 0.4$ kpc.
   We further cross check this estimated distance by using the color-magnitude diagrams (CMDs) \citep{2022ApJ...926...25P, 2020ApJ...891...81P, 2020MNRAS.498.2309S, 1994ApJS...90...31P}. 
    In the right panel of Figure~\ref{cc}, we have shown the $V$ versus 
    $(V-I)$ CMD for the stars in the cluster region. The member and massive
    stars are also marked in the figure. The curves represent the ZAMS from \citet{2019MNRAS.485.5666P} corrected for a distance of $3.8$ kpc, and reddening $E(B-V) \sim0.45$ mag (blue curve) and $E(B-V) \sim0.80$ mag (red dotted curve), for the foreground and cluster reddening values, respectively.  Clearly, the massive stars are falling correctly on the respective massive end of the ZAMS for a distance of 3.8 kpc and  $E(B-V) = 0.80$ mag.
    We have also used PanSTARR1 optical data to substantiate our parameters. For this purpose, 
    we have plotted the $g$ versus $(g-i)$ CMD for the stars located inside the cluster region (black dots), massive stars (star symbols, and the identified cluster members (green circles) in Figure~\ref{cmd}.
    In left panel of Figure~\ref{cmd}, the ZAMS \citep{2019MNRAS.485.5666P} corrected for different  extinction and distance values (as reported in literature) are shown and in the right panel the present estimates are used. 
    Certainly, out of different distance and reddening estimates, the ZAMS corrected for the present estimates (distance $\sim3.8$ kpc and $E(B-V)\sim0.80$ mag) seems to be a best match to the distribution of stars within the Boc2 cluster.
    
\begin{figure}
    \centering\includegraphics[height=0.26\textheight]{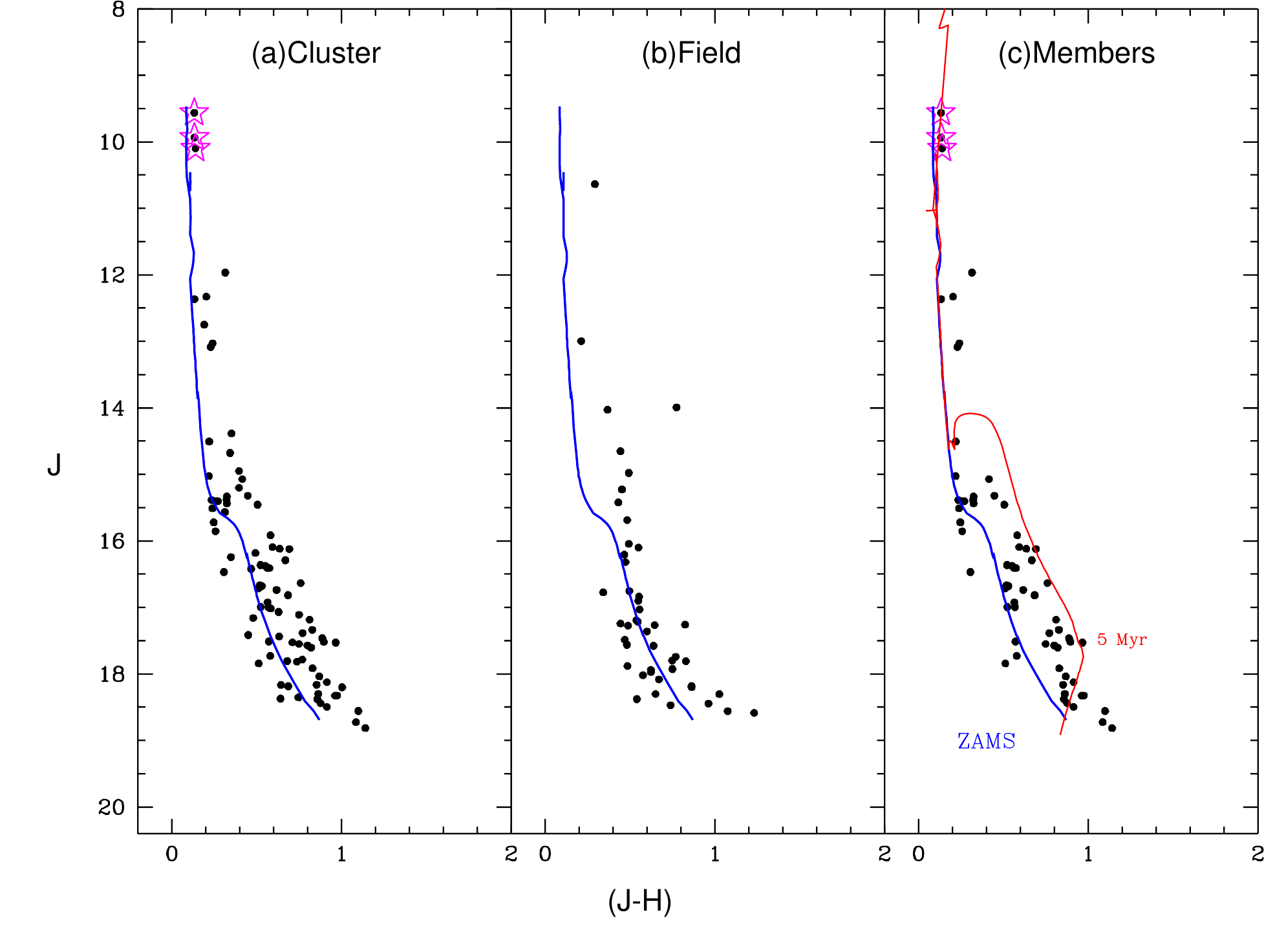}   
    \caption{$J$ versus $(J-H) $ CMD for (a) stars within the cluster region, (b) stars within the reference field of same area as of the cluster  and (c) statistically cleaned sample, which is over-plotted with the  ZAMS (solid blue curve) and the 5 Myr isochrone (solid red curve) taken from \citet{2019MNRAS.485.5666P}. All the isochrones have been corrected for a distance of $3.8$ kpc and reddening $E(B-V)=0.80$ mag. Location of massive stars are shown by the magenta star symbols.} \label{scmd}
    \end{figure}

   As already discussed, the previous studies suggests that, this cluster is young (from 4.6 to 7 Myr) in nature \citep{1979A&AS...38..197M, 1993AJ....105.1831T, 2015MNRAS.450.4150C, 1995MNRAS.277.1269M, 2016A&A...585A.101K}.
   Most of these estimated ages were based on the MS lifetime of the most massive stars in the cluster. In Figure \ref{scmd}, we show the $J$ versus $(J-H) $ CMD for (a) stars within cluster region, (b) stars within reference region (or field region) of same area as of cluster and (c) statistically cleaned sample of stars \citep[see also;][]{2007MNRAS.380.1141S,
   2012PASJ...64..107S,2017MNRAS.467.2943S, 2008MNRAS.383.1241P, 2013ApJ...764..172P,2011MNRAS.415.1202C,2013MNRAS.432.3445J}. We have also marked the location of massive stars in the CMDs. The clusters CMD distribution matches well with the ZAMS, specially in the lower part of the CMD. Even after the statistical subtraction, the distribution of stars in the lower part of the CMDs stays similar. We have plotted the isochrone of $5$ Myr (approx age of massive stars) over the statistically cleaned CMD.  The upper end of the CMD (location of massive stars) is matching well with the $5$ Myr isochrone but in the lower-part of the CMD, there are not many stars falling on the $5$ Myr isochrone. This kind of distribution suggests an extended  history of star formation in this cluster, i.e., the star formation in Boc2 cluster might have started much earlier than the formation of massive stars \citep[cf.][]{2005MNRAS.358.1290P}.
    From the current distribution of stars in the CMDs, its very difficult to constrain the first epoch of star formation in the Boc2 cluster, but the most recent star formation have occurred at 5 Myr ago with the formation of massive stars.

\subsection{Mass Function}
\begin{figure}
    \centering\includegraphics[height=0.34\textheight]{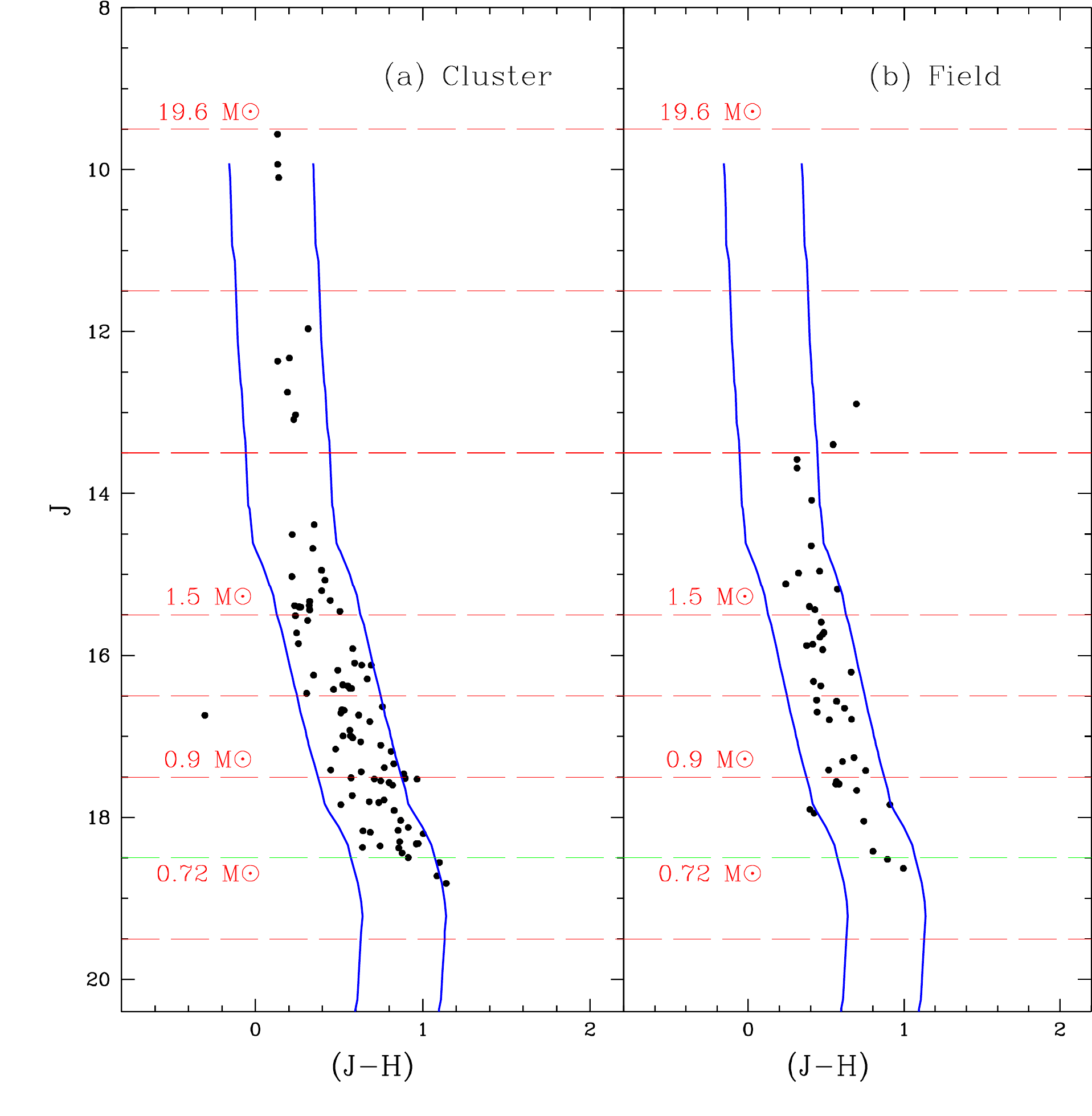}
    \caption{$J$ versus $(J-H) $ CMD for (a) stars within the cluster region, 
    (b) stars within the field or reference region of the same area as of cluster.
    The curves denotes  envelop of +0.3 mag (right curve;  $A_{v} = 2.42$ mag) and -0.2 mag (left curve; $A_{v}=1.88$ mag) around the CMD from the isochrone of \cite{2013ApJS..208....9P}. Green horizontal line shows the magnitude limit corresponding to the completeness factor of 80\%}\label{lf}
    \end{figure}

    The initial mass function (IMF) is known as the distribution of 
    stellar masses that form in one star formation event in a given volume of space. It is one of the important statistical method to study star 
    formation. Open clusters possess many favorable characteristics 
    for MF studies, e.g., clusters containing almost  the coeval set of stars at the same distance with the same metallicity; hence, difficulties such as complex corrections for stellar birth rates, life times, etc, associated with determining the MF from field stars are automatically removed. The MF is often expressed by a power law, 
    N(log m)$\propto$ m$^\Gamma$ , and the slope of the MF is given as:
    
    \begin{equation}
    \Gamma = {{d logN(logm)}\over{d logm}}
    \end{equation}
    
    where $Nlog(m)$ is the number of stars per unit logarithmic mass interval. The luminosity function (LF) of the cluster is converted into MF using a theoretical evolutionary track \citep[][]{2013ApJS..208....9P}.  
    To obtained the LF, we generate $J$ versus $(J-H)$ CMD from the NIR data (cf. Figure~\ref{lf}), corrected for the data incompleteness, distance and extinction \citep[see][]{2020MNRAS.498.2309S}.
    In the Figure~\ref{lf}, we show the CMD for the
    cluster region as well as for the field region ($\alpha_{J2000}$: 06$^h$49$^m$08.67$^s$, $\delta_{J2000}$: 
    	+00$^\circ$27$^\prime$32.36$^\prime$$^\prime$), having
    the same area. The contamination due to field
    stars is greatly reduced by selecting a sample of
    stars which are located near the well-defined main sequence (MS) \citep[cf.][]{2020MNRAS.498.2309S,2008AJ....135.1934S}. Thus, we have generated an envelope of +0.3 mag and -0.2 mag around the CMD considering 
    the distribution of member stars and is shown in the left panel of Figure~\ref{lf}. The number of probable cluster members were obtained 
    by subtracting the contribution of field stars (corrected for data incompleteness), in different magnitude bins,
    from the contaminated sample of cluster stars (corrected for data incompleteness).
        
    The photometric data can be incomplete due to various reasons, e.g., nebulosity, crowding of the stars, detection limit, etc. 
    We estimated the completeness of NIR data by using the distribution of star number in different mag bins as shown in the left
    panel of Figure~\ref{mf}. 
    The peak in the stellar number distribution will more or less represent the complete sample after which the completeness of the photometric data tends to decrease \citep[see also,][]{2020MNRAS.498.2309S, 2017ApJ...836...98J}.
    Thus, we have found that the present NIR data (2MASS+UKIDSS) is complete
    upto $\sim17.5$ mag in J band, corresponding to a star of $\sim0.9$ M$\odot$ at a distance of 3.8 kpc. 
    For the stars in mag bin $17.5-18.5$ in J band, we have used the completeness factor of 80\%  \citep[see also,][]{2020MNRAS.498.2309S,2017ApJ...836...98J}.
    The resultant MF distribution in the
    cluster region is shown in right panel of Figure~\ref{mf}.

    \section{Discussion}
    
\begin{figure*}
\centering\includegraphics[width=0.49\textwidth]{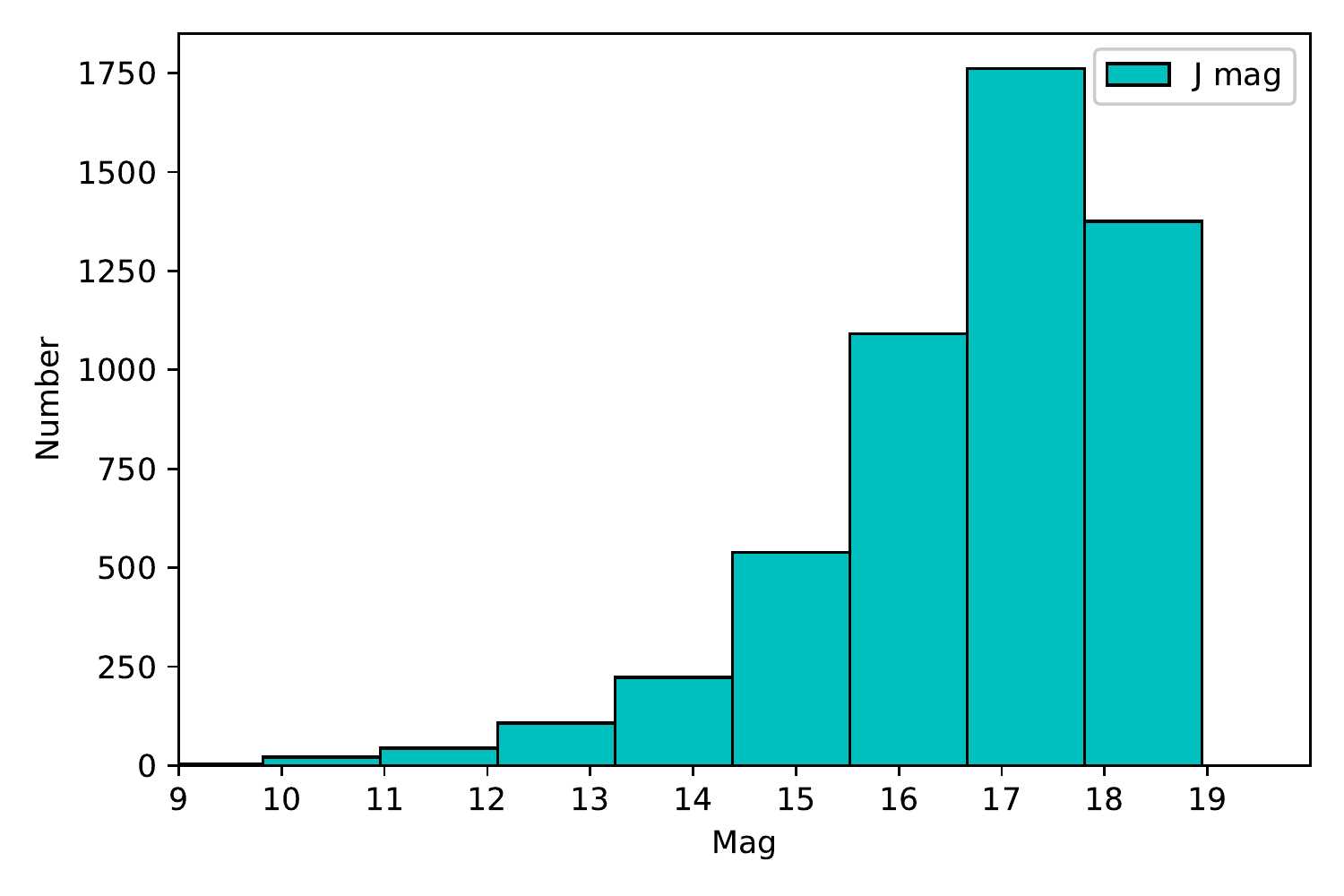}
\centering\includegraphics[width=0.49\textwidth]{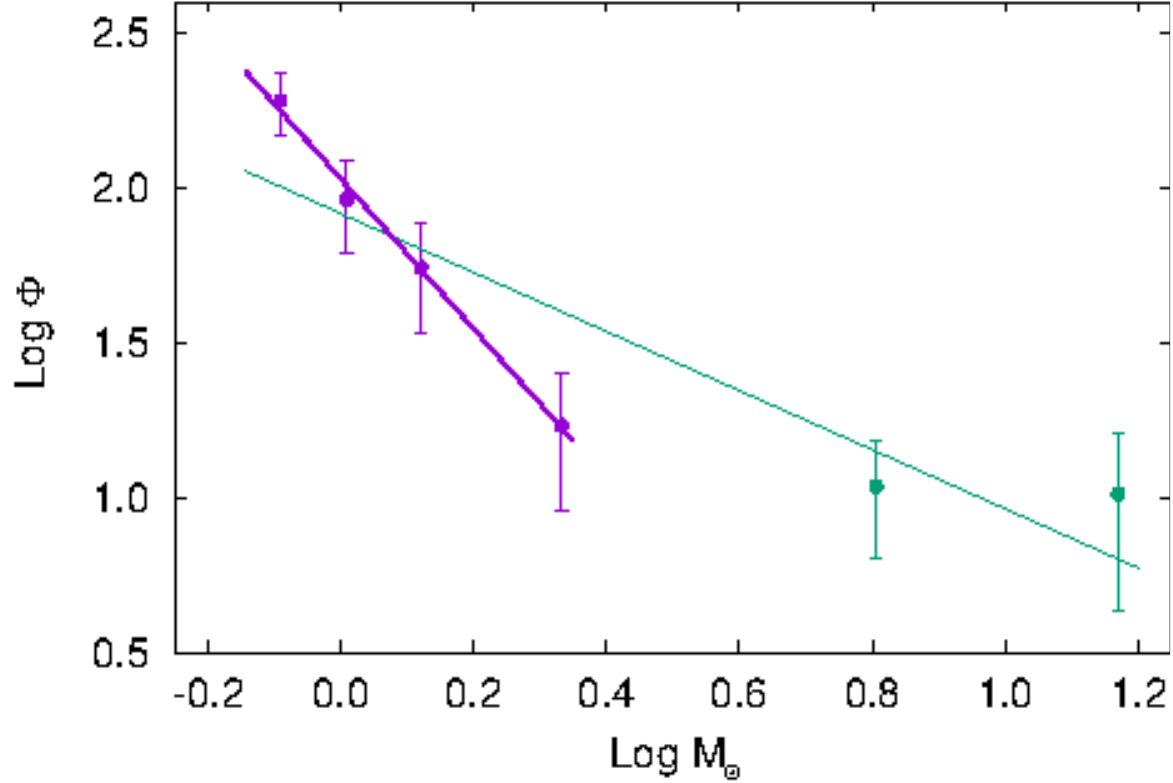}
\caption{{\it Left panel}: The histogram shows the data completeness of J mag of NIR data. {\it Right panel}: A plot of the mass function (MF)  for the Boc2 cluster using NIR data. Log$\phi$ represent log(N/dlog m) and the error bars represents standard $\pm\sqrt{N}$ errors. The solid line shows a least squares fit to the MF distribution.}\label{mf}
\end{figure*}

    It is known that the higher mass stars mostly follow the 
    Salpeter MF \citep[$\Gamma=-1.35$;][]{1955ApJ...121..161S}.
    At lower masses, the MF is less  constrained, but appears to
    flatten below 1 M$_{\odot}$ and exhibits few stars of the lowest masses
    \citep{2002MNRAS.336.1188K, 2003PASP..115..763C, 2016ApJ...827...52L}.
    In this study, we find that the MF distribution for the lower mass bins ($\sim$0.72$<$M/M$_{\odot}$$<$2.8) is showing a well defined linear distribution (cf. left panel of Figure~\ref{mf}), and has a   
    slope ($-2.42\pm0.13$) which is very much steeper than the Salpeter value ($-1.35$). This indicate towards the presence of the excess number of low-mass stars in the Boc2 cluster.
    This apparent excess of low mass stars could be due to the stellar evolutionary effects in which massive stars could have evolved off from the MS. 
    If we consider massive stars in the distribution ($\sim$0.72$<$M/M$_{\odot}$$<$19.6), the value $\Gamma$ becomes shallow ($-0.95\pm0.24$, cf. left panel of Figure~\ref{mf}), indicating the formation of massive stars in the recent epoch of star formation. The age analysis in Section 3.3 also suggest similar results in which low mass stars in the Boc2 cluster which might have formed at an earlier epoch of star formation than the massive ones.

      To investigate further, we have calculated the mass segregation ratio (MSR) as a measure to quantify mass segregation in the cluster region by using  \citet{2009MNRAS.395.1449A} method. In this approach, we constructed the
    minimum sampling tree (MST) for the massive stars and for the equal
    number of randomly selected stars from the cluster sample and estimated  their mean edge length as $\gamma_{mp}$ and $\gamma_{rand}$ respectively. 
    Therefore, the value of $\Gamma_{MSR}$ \citep[cf.][]{2020MNRAS.498.2309S, 2020yCat..74730849D, 2011A&A...532A.119O} is estimated as: 
      \begin{equation}
        \Gamma_{MSR} = {\langle{\gamma_{MST}^{rand}}\rangle\over{\gamma_{MST}^{mp}}}
      \end{equation}
      A value of $\Gamma_{MSR} \sim1$ implies that both most massive and randomly selected samples of stars are distributed in a similar manner, whereas $\Gamma_{MSR} > 1$ indicates mass segregation and $\Gamma_{MSR} << 1$ points to inverse mass segregation \citep{2020yCat..74730849D}.
      To minimize the uncertainties associated with the conversion of luminosity to mass, we have used magnitude of cluster stars (cf. Figure~\ref{lf}) as a proxy to their masses \citep[cf.][]{2020yCat..74730849D}.
      For the Boc2 cluster, we derived $\Gamma_{MSR} = 1.3 \pm 0.9$, which 
      suggests the effect of mass segregation in this cluster.
      To further check, whether mass segregation is primordial or due to 
      dynamical relaxation, we estimated the dynamical relaxation time,
      $T_E$, the time in which the individual stars exchange sufficient energy
      so that their velocity distribution approaches that of a Maxwellian
    equilibrium using the method given by \citet{1987gady.book.....B}:
      \begin{equation}
        T_E = {N\over{8logN}}\times T_{cross}
      \end{equation}
     Where, $T_{cross}$=$D/\sigma_v$ denotes the crossing time, $N$ is the total number of stars in the cluster region under study of diameter D, and 
     $\sigma_v$ is the velocity dispersion, with a typical value of 
     3 km s$^{-1}$ \citep{2017NewA...52...55B}. For the Boc2 
     cluster, the value of dynamical relaxation time comes
     out to be $T_E \sim8$ Myr \citep[cf.][]{2020MNRAS.498.2309S}.
     If we assume the loss of $50\%$ of stars due to incompleteness of our data, the dynamical relaxation time will be $T_E \sim14$ Myr, which is
    more than the age of the massive stars in the Boc2 cluster(i.e., $\sim5$ Myr). This suggest that massive stars in the Boc2 cluster have formed in the inner regions of this cluster.

       As the cluster becomes older, the stars escape from the cluster region because of various processes. These can be divided
    into internal and external processes. Internal processes are like two body
    relaxation and dynamical relaxation \citep{2015ApJ...810...40D}, and external
    processes are like tidal interaction and encounters with molecular clouds,
    spiral arm and Galactic disc \citep{2006MNRAS.371..793G,2008MNRAS.389L..28G, 2020A&A...635A.125Y}. 
   Since, the Boc2 cluster hosts older generation of low mass stars, there is also a possibility that some of them escaping from the cluster region. Thus, we have calculated the tidal radius of this star cluster by using the equation given by \citet{1998MNRAS.299..955P}:
    
    \begin{equation}
    r_{t} = \left[{{GM_C}\over{2(A-B)^2)}}\right]^{1/3} 
    \end{equation}
    
     where $G$ is the gravitational constant, $M_C$ is the total mass of the
    cluster, and $A$ and $B$ are the Oort constants, $A = 15.3 \pm 0.4$
    km s$^{-1}$ kpc$^{-1}$, $B = -11.9 \pm 0.4$ km s$^{-1}$ kpc$^{-1}$ \citep{2017MNRAS.468L..63B}. From the conversion of LF into 
    MF (cf. Section 3.4), mass of the Boc2 cluster is estimated as $M_c =126$ M$_{\odot}$. The corresponding tidal radius comes out to be $\sim$7 pc. This is a good approximation even if we have
    missed 50 per cent of the cluster mass in the lower mass bins due to
    data incompleteness, then the resultant tidal radius will be $\sim$9 pc. Therefore, we can conclude that the tidal radius ($r_t$ $\sim$ 7 pc) of this cluster is much larger than the present estimate of the cluster radius ($R_{cluster}$ $\sim$1.1 pc).
    Hence, the observed small size of this cluster resembles a remain of an old population of stars formed via a earlier epoch of star formation. A recent epoch of the star formation activity in this cluster might be responsible for the formation of massive stars in this cluster.

\section{Summary and Conclusion}

    We have performed a deep (V$\sim$21.3 mag) and wide-ﬁeld  (FOV $\sim18.5\times18.5$ arcmin$^{2}$) multiband ($UBVI_{c}$) photometric  observations around the Boc2 cluster. 
    The optical data, along with Gaia DR3 and deep PanSTARR1 (PS2) data have been used to study the membership probability of stars in the 
    cluster region, structural parameters of the cluster, MF and mass segregation in the Boc2 cluster. The main results are summarized as follows:
    \begin{itemize}
    
    \item We have derived the structural parameters of the Boc2 cluster by using isodensity contour (NN method) analysis and found that this cluster shows circular morphology. The cluster is having radius of 1 arcmin (1.1 pc) and is centered at  $\alpha_{J2000}$: 06$^h$48$^m$50$^s$, $\delta_{J2000}$: +00$^\circ$22$^\prime$43.611$^\prime$$^\prime$.

     \item Using Gaia DR3 proper motion data, we have identified 24 stars as most 
     probable cluster members of the cluster Boc2. The distance of this cluster, as estimated from the Gaia parallax and isochrone fitting method, comes out to be $\sim3.8\pm0.4$ kpc. 
     The reddening towards the cluster is estimated as $E(B-V)\sim0.8$ mag. 
     We also estimated age of the massive candidates of this cluster as $\sim5$ Myr. This cluster also seems to hold an older population of low mass stars.
     
     \item By using  deep NIR data, we also derived MF slope
     ($\Gamma$) in the cluster region as 
     $-2.42\pm0.13$ in the mass range $\sim$0.72$<$M/M$_{\odot}$$<$2.8. This slope is steeper then Salpeter
     value ($-1.35$). This indicates the presence of a excess number of
     low-mass stars in the cluster.

    \item This cluster shows the effect of mass segregation whereas the dynamical age ($\sim8-14$ Myr) of this cluster is found to be more than the age of massive stars ($\sim5$ Myr). This indicate that the massive stars have formed in the inner region of the Boc2 cluster, in a recent epoch of star formation.

    \item The tidal radius of the Boc2  cluster  ($\sim7-9$) is much more than its observed radius ($\sim1.1$ pc). This indicate that the most of the stars in this cluster are the remnants of an older population of stars formed via an earlier epoch of star formation.

     \end{itemize}
\section*{Acknowledgements}
\vspace{1em}
The observations reported in this paper were obtained using the 1.3m Devesthal Fast 
Optical Telescope, Nainital, India. This work is based on data obtained as part of
the UKIRT Infrared Deep Sky Survey (UKIDSS). This publication made use of data
products from 2MASS (a joint project of the University of Massachusetts and the 
Infrared Processing and Analysis Center/ California Institute of Technology, funded
by NASA and NSF) and archival data obtained with the Spitzer Space Telescope 
(operated by the Jet Propulsion Laboratory, California Institute of Technology,
under a contract with NASA). This study has made use of data from the European
Space Agency (ESA) mission Gaia (https://cosmos.esa.int/ gaia), processed by the
Gaia Data Processing and Analysis 
Consortium (DPAC; https://cosmos.esa.int/web/gaia/dpac/consortium). 
Funding for the DPAC has been provided by the institutions participating in 
the Gaia Multilateral Agreement. 

\bibliography{boc2}




\end{document}